\documentclass{elsart}
\usepackage{graphics}

\begin{document}
\begin{frontmatter}

\title{Stimulation of Beta Decay due to a Bose-Einstein Condensate}
\author{J.J. Hope and C.M. Savage}
\address{Department of Physics and Theoretical Physics, \\
The Australian National University,
Australian Capital Territory 0200, Australia. \\
Joseph.Hope@anu.edu.au}

\begin{abstract}
Nuclear processes can be stimulated by the presence of a macroscopic number 
of bosons in one of the final states.  We describe the conditions necessary 
to observe the atom-stimulation of a beta decay process.  The stimulation 
may be observable if it becomes possible to produce a Bose-Einstein 
condensate with the order of $10^{14}$ atoms in a trap.
\end{abstract}


\end{frontmatter}

It has recently been noted that the rate of gamma ray emission from a 
nuclear process can be altered by the presence of a Bose-Einstein 
condensate (BEC) \cite{Namiot}.  This process of \emph{atom}-stimulated 
photon emission is similar to other proposals involving electronic 
transitions \cite{Javanainen,You94,You96,Hope}, except that it emphasises 
the fact that any process, including nuclear transtions, may be enhanced.  
We calculate the stimulation of the emission of a \emph{massive} particle, 
in particular a beta particle, and discover that the necessary 
experimental conditions are less restrictive.  The origin of this 
improvement is that the total momentum kick due to the neutrino and 
the beta particle may be zero.

A large number of atoms in a single quantum mechanical (internal and 
external) state has been produced in atomic traps in the recent 
experiments which have produced a Bose-Einstein condensate 
\cite{Rb,Li,Na}.  The high degree of quantum degeneracy in these 
systems may be used to stimulate any process which has a final product 
which is condensed.  Any transition rate between two states is 
enhanced by a factor of $(N+1)$ where $N$ is the number of bosons 
occupying the final condensed state.  One of the interesting features 
of condensed matter physics is that under certain conditions, large 
collections of an even number of bound fermions (i.e.  atoms) can act 
as bosons.  This approximation is true when the Hamiltonian describing 
the fields may be written only in terms of field operators for the 
entire atoms, which will hold for nuclear reactions when there is 
negligible probability of atomic disruption.  It would be extremely 
interesting to test this fundamental assumption of condensed matter 
physics with such an exotic example.  The stimulation of any kind of 
transition by a BEC has not been observed, except indirectly, in the 
formation of the condensate itself.

Recently, there have been several similar proposals to measure the effect 
of the Bose enhancement of photon emission in atomic systems using a BEC. 
These include the work on light scattering from a BEC 
\cite{Javanainen,You94,You96}, the stimulated enhancement of cross section 
(SECS) for excited state atoms moving through a BEC\cite{Hope}, and the 
emission of a gamma ray photon from a nuclear transition \cite{Namiot}.  It 
is possible that the light scattering proposal and the SECS proposal may be 
realised with current condensates.  The gamma ray proposal was found to 
require the construction of a condensate with $10^{14}$ atoms and have a 
wavefunction which was a metre long in one dimension.  It also involved two 
of these condensates fired at each other with a very well defined relative 
momentum which was equal to the recoil kick of a gamma ray photon.  This 
appears to be a difficult experimental task, and there is a strong 
likelihood that at such energies there would be sufficient coupling between 
the two condensates to destroy them.  Atom-stimulation of a photon emission 
is also used to produce a build up of atoms in the ground state of a trap 
in some of the recent models for an atom laser \cite{Wiseman,Spreeuw,Moy}.  
The atom laser proposals have yet to be realised.

We analyse a different system in which the stimulation of a 
\emph{massive} particle can be detected.  We consider a radioactively 
unstable ion, $A$, that is held in a trap where it will decay into the 
stable atom, $B$, as well as a beta particle and an (anti)neutrino.

\begin{equation}
	A \rightarrow B + \beta + \bar{\nu}
	\label{NuclearEqn}
\end{equation}
The unstable particle $A$ must be an ion so that after the beta decay 
it will be a stable atom of the new species without having to capture 
or emit an electron.  If there is a second trap which contains a 
condensate of $N$ atoms of $B$, then there will be an enhancement of 
the fraction of the nuclear decay which goes into the ground state of 
the trap, and therefore an enhancement of the overall decay rate.  It 
is important to note that only a very small number of unstable ions of 
species $A$ are needed in the first trap.  We let $\gamma$ be the 
spontaneous nuclear decay rate and $f$ be the fraction of the atoms 
which will be in the ground state of the trap after a spontaneous 
decay.  The total decay rate $\gamma_{tot}$ is the sum of the 
spontaneous decay rate into the non-ground states and the stimulated 
decay into the ground state

\begin{equation}
	\gamma_{tot} = \gamma (1-f) +(N+1)\gamma f = \gamma(1+Nf).
	\label{EmissionRate}
\end{equation}

It is clear from this equation that for a BEC with a sufficiently large 
number of atoms, the decay rate can be detectably increased, and that the 
fraction $f$ will determine the critical size of such a BEC.  We will now 
calculate this fraction for reasonable parameters.

We denote the center of mass wavefunction of the unstable ion $A$ by 
$\Psi({\bf x})$.  We also denote the mean field of $B$ when it is in the 
condensate by $\Phi({\bf x})$, and the net momentum kick produced by the 
beta particle and neutrino emission by $\hbar {\bf k}$.  The fraction 
$f$ of atoms which would decay spontaneously into the ground state of the 
trap for $B$ is given by the overlap integral

\begin{equation}
	f = 1/L \int d^3{\bf k} \: g({\bf k}) \left\|\int d^3{
	\bf x}\, \Phi^*({\bf x})\Psi({\bf x}) e^{i{\bf k}.{\bf 
	x}}\right\|^{2},
	\label{eq:f}
\end{equation}
where $g({\bf k})d^3{\bf k}$ is the probability of the momentum kick being 
between ${\bf k}$ and ${\bf k}+d^3{\bf k}$, and $L$ is the number of 
magnetic sublevels in the nucleus.  The fraction $f$ is reduced by this 
factor because the beta decay will randomise the nuclear magnetic moment.  

The beta particle will be emitted at relativistic speeds, and will 
therefore have virtually no interaction with the orbital electrons.  
This is due to an extremely poor spatial overlap between the 
wavefunction of the outgoing beta particle and the bound electrons.

The momentum kick distribution $g({\bf k})$ can be found from the dynamics 
of the reaction and the momentum distribution of the emitted particles.  
Conservation of energy and momentum allow us to determine the magnitude of 
the kick $k(p,\theta) = |{\bf k}|$ given to the atom:

\begin{eqnarray}
	\nonumber
	k(p,\theta) &=& \frac{1}{\hbar}\left[\frac{\Delta\/E^2}{c^2} +2 p^2 +m^2c^2
	 +2p \cos(\theta) \frac{\Delta\/E}{c} \right.\\
	&&\left.\rule{30mm}{0mm} -2(\Delta\/E/c+p \cos(\theta))\sqrt{p^2+m^2c^2
	}\right]^{1/2}
	\label{eq:kick}
\end{eqnarray}
where $m$ is the mass of the beta particle, $p$ is the momentum of the 
beta particle, $\Delta\/E=Q+mc^2$ is energy difference between the 
nuclear states of $A$ and $B$, and $\theta$ is the angle between the 
beta particle and the neutrino.

The distribution of the momentum kick is isotropic, and it can be 
calulated from the momentum distribution of the beta particles:

\begin{equation}
	g(k) = \int_0^{p_{max}} dp \int_0^\pi \;d\theta \;\delta \left( k(
	p,\theta)-k \right) \;G(p)
	\label{eq:gGlink}
\end{equation}
where $p_{max}=\sqrt{\Delta\/E^2/c^2 -m^2c^2}$, and the function $G(p)$ is 
the momentum distribution of the beta particles.  This distribution is 
found from the density of states \cite{Frauenfelder}, and is given by $G(p) 
= p^2 (\Delta\/E - \sqrt{p^2 +m^2c^4})^2/\mathcal{N}$, where $\mathcal{N}$ 
is simply a normalization factor.  We can calculate this factor to be

\begin{eqnarray}
    \nonumber 
	\mathcal{N}&=&4/5 \;c^2 \;(\Delta\/E^2/c^2-m^2c^2)^{5/2} +\\
    \nonumber
    &&\rule{10mm}{0mm}\Delta\/E^2/c^2 \;\sqrt{\Delta\/E^2/c^2-m^2c^2} 
    \;(m^2c^4-2\Delta\/E^2) + \\
    \nonumber
    &&\rule{10mm}{0mm}4/3 \;(\Delta\/E^2/c^2-m^2c^2)^{3/2}\;(\Delta\/E^2
    +m^2c^4) +\\
    \nonumber
    &&\rule{10mm}{0mm}c^5 \:m^4 \Delta\/E \;\ln{\frac{\Delta\/E 
    +\sqrt{\Delta\/E^2-m^2c^4}}{mc^2}}.
\end{eqnarray}

We now consider a specific case to calculate the overlap integral, 
Eq.\ (\ref{eq:f}).  We consider the case where both wavefunctions have 
a Gaussian form, which corresponds to the ground state of an atom trap 
containing non-interacting atoms.  For simplicity, we assume that both 
wavefunctions are isotropic and identical.

\begin{equation}  \label{eq:gstate}
	\Psi({\bf x}) = \Phi({\bf x}) = (\frac{1}{2 \pi l^2})^{\frac{1}{4}} 
	\mbox{exp}\left[-\frac{{\bf x}^{2}}{4l^{2}}\right],
\end{equation}
where $l$ is related the size of ground state of the trap.  We have 
assumed here that the traps containing the two species have been 
perfectly aligned.

The traps which have been used in BEC experiments so far have been 
magnetic, which could trap the two prepared species of atoms, although 
in general the size of the ground state of the trap would be different 
for each species.  In an experiment, the traps will have to be chosen 
so that the shape of the wavefunctions of the two states are as 
similar as possible.  This could be possible with dipole force traps 
\cite{Aminoff}, which could also be used to prepare the atoms $A$ and 
$B$ separately and then move them together by slowly moving the traps.  
The disadvantage with optical traps is that they may excite the atoms, 
causing losses through spontaneous emission.  This may be avoided 
through using extremely large detunings or Raman transitions 
\cite{Hope96}.

From Eqs.\ (\ref{eq:f},\ref{eq:kick},\ref{eq:gGlink},\ref{eq:gstate}) we 
calculate the fraction $f$ of atoms spontaneously decaying into the 
ground state under our assumptions,  

\begin{equation}  
\label{eq:fresult}
	f = \frac{\pi^2 \hbar^2}{8c} \frac{(\Delta\/E^2-m^2c^4)^3
	          (\Delta\/E^2+m^2c^4)}{L\; \Delta\/E^5\; l^2 \; \mathcal{N}}.
\end{equation}

This decreases with increasing $\Delta\/E$, which means that the best 
reaction will have a low maximum kinetic energy $Q$ for the emitted 
beta particle.  Examination of the dependence of $f$ on the mass $m$ 
of the emitted particle shows that beta decay will be stimulated more 
than a different two body decay involving a heavier particle.  From 
the inverse square dependence on $l$, we see that a small ground state 
wavefunction will also increase $f$.  The size of the ground state 
wavefunction is determined by the strength of the trap.  The experiments 
which currently have produced a BEC have had traps corresponding to 
$l\approx 2-6 \mu$m.

To obtain a reasonable signal from this experiment, it is necessary to 
have enough ions trapped so that a useful number of decays happen 
over the lifetime of the experiment.  Since they must be trapped 
within such a small region, it will be difficult to get more than a 
thousand ions into the trap.  The stimulation will also require a 
large number of atoms in the condensate.  The density of atoms in the 
condensate cannot be increased indefinitely, however, as collisions 
and interatomic attraction will eventually destroy the condensate.  
The mean interatomic spacing should be larger than the scattering 
length of the atoms, which imposes a condition on $l$, the size of the 
ground state wavefunction.  The required number of atoms varies as 
$l^2$, so the density will go down if larger wavefunctions are used.  
There is a possibility that the scattering length can be tuned with an 
applied magnetic field, which is true for cesium \cite{Tiesinga}.  If 
this technique can be applied more generally, it would allow quite 
high atomic densities in the trap.

In the presence of a strong bias magnetic field, the electronic and 
nuclear spins are not coupled.  The magnetic moment of the nucleus 
will be randomised, so the number $L$ of internal states available to 
the product atom $B$ will be $2I+1$, where $I$ is the spin of the 
final nuclear state.  We use this assumption when calculating the 
results in Table $1$, in which we show the number of atoms required to 
double the expected rate of beta decay for several different nuclei.  
This table uses the parameter $l=2\mu$m for the size of the trap.  The 
reactions on this table were chosen so that less than a thousand ions 
were required in the trap to produce a sufficiently large signal.  
This restriction excluded several reactions which used atomic species 
which have already been trapped, and shows that the best results are 
found using species which have not yet been cooled below recoil limited 
temperatures.  

\begin{table}
	\caption{Comparison of the stimulation of beta decay from 
	different nuclei}
	\protect
        \label{tab:comp}
\begin{tabular}{|c|c|c|c|c|c|} 
\hline
Initial&Final&Half life&$Q$&$N$ required
                &Ions needed to\\
nucleus&nucleus&(min)&(MeV)&to double&produce 1 decay/min\\
&&&&decay rate&($N=10^{14}$)\\
\hline 
$^{42}$K&$^{42}$Ca (I=0)&$744$&$3.5211$&$1.7\times 10^{14}$&$670$\\
\hline 
$^{11}$C&$^{11}$B (I=3/2)&$20.4$&$1.9821$&$2.4\times 10^{14}$&$21$\\
\hline
$^{31}$Si&$^{31}$P (I=1/2)&$157$&$1.4908$&$7.4\times 10^{13}$&$96$\\
\hline
\end{tabular}
\end{table}

The first reaction on this table was the $\beta^{-}$ decay of $^{42}$K 
to $^{42}$Ca, for which the decay rate will be doubled if there are 
$1.7\times 10^{14}$ atoms in the BEC. If there are $10^{14}$ atoms in 
the BEC, then there will be a $60\%$ increase in the decay rate, and 
this new rate will be one decay per minute when there are $670$ ions 
in the trap.  These results require extremely large condensates of 
exotic species, so it is unlikely that the stimulation of a nuclear 
process will be observed in the near future.  If it becomes possible 
to produce a BEC with $10^{14}$ atoms, this feasibility study shows 
that it may indeed be possible to measure the stimulation of a beta 
decay process.  The largest BEC produced so far has contained $5\times 
10^{6}$ atoms \cite{largeBEC}, which is more than two orders of 
magnitude larger than the first BEC produced less than a year earlier 
\cite{Rb}.  

The analytic form, Eq. (\ref{eq:fresult}), of the result was found by 
choosing an explicit and simple form for the shape of the wavefunctions.  
This means that the quantitative results for particular experiments may 
vary, but this is not important for a proof of principle calculation.  In 
particular, the ground state of a harmonic trap will be altered by the 
presence of interatomic interactions, which are modelled by the non-linear 
Schr\"{o}dinger equation\cite{NLSE}.  The maximum overlap will be obtained 
when the wavefunction of the unstable atoms is matched to the ground state 
wavefunction of the BEC. For repulsive interactions, this wavefunction will 
be spread slightly (for an example, see the calculation in reference 
\cite{You96}).

This letter has examined the feasibility of measuring a stimulation of 
beta decay by a large number of bosonic atoms in a single quantum 
mechanical state.  It shows that while there is no fundamental 
obstacle to such an experiment, the required condensate population is 
eight orders of magnitude beyond that achieved so far.

\section*{Acknowledgement}

The authors would like to thank A. White and A. Baxter for their 
stimulating and helpful discussions.

\end{document}